# Laser-induced fluorescence quenching of red fluorescent dyes with green excitation: avoiding artifacts in PIE-FRET and FCCS analysis

Mikhail Baibakov and Jérôme Wenger

*Aix Marseille Univ, CNRS, Centrale Marseille, Institut Fresnel, Marseille, France*

Email : jerome.wenger@fresnel.fr

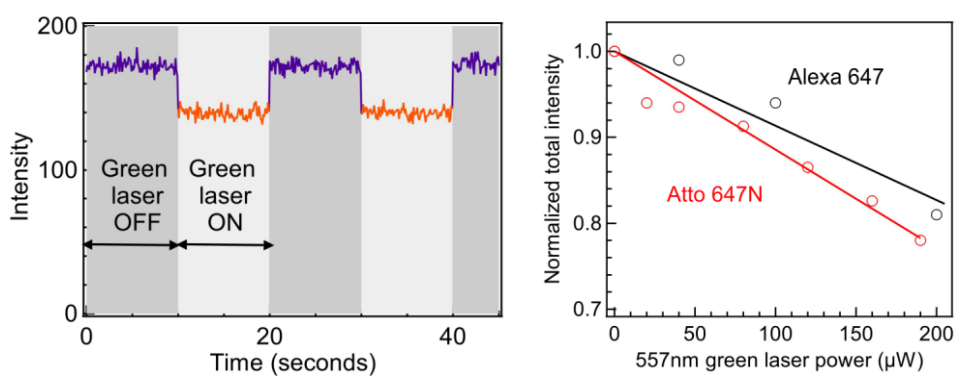

**Abstract :**

Multicolor excitation is at the core of many fluorescence spectroscopy techniques such as PIE-FRET, ALEX-FRET, or FCCS. However, the influence of the multiple laser excitations on the dye photophysics is often overlooked. Here, we show that green laser pulses can surprisingly quench the fluorescence of common red dyes Alexa Fluor 647 and Atto 647N, even when the conditions leading to photobleaching are avoided. The physical origin of this phenomenon is discussed via a long-lived dark state and/or photorefractive effects. These observations are important to avoid measurement artifacts as both the fluorophore concentration and the fluorescence brightness are affected.

1. Introduction

Multi-color excitation and detection of fluorescent species is gaining a large interest for confocal biological imaging [1–3] and single molecule studies [4,5]. In contrast to their single-color counterpart, multi-color approaches add another dimension to monitor molecular interactions, enzymatic reactions or dynamical structural changes of complex macromolecules. Among these techniques, Förster resonance energy transfer (FRET) and fluorescence cross-correlation spectroscopy (FCCS) are currently the most widely used [6]. FRET describes the energy transfer from an excited donor fluorophore to an acceptor, and is a well-established method for determining distances between the labeled sites over 2-10 nm range [7]. FRET is therefore particularly attractive for monitoring dynamic structural changes of proteins and complex biomolecules. Alternatively, FCCS exploits the fluorescence intensity fluctuations induced by molecular diffusion accross the confocal detection volume to monitor colocalization and interaction between different fluorescent dyes [8]. FCCS therefore provides molecular information complementary to FRET, and can be used for applications where donor-acceptor separations are larger than 10 nm [9].

A prevalent issue in FRET and FCCS is related to incomplete labelling, e.g. donor-labeled structures lacking an acceptor, or acceptor-labeled structures lacking a donor [10]. Therefore, techniques involving the use of two alternating laser excitations have been introduced to excite respectively the donor and the acceptor fluorescence [11,12]. These techniques are known as alternating laser excitation (ALEX-FRET) and pulsed interleaved excitation (PIE-FRET). In ALEX-FRET, the two laser beams are interleaved on a microsecond timescale [11], while in PIE-FRET, picosecond lasers are interleaved on a nanosecond timescale [12].

A common feature in ALEX-FRET, PIE-FRET and FCCS is the use of multiple laser beams to excite the different fluorophores. However, the mutual influence of these different excitation pathways on the dye photophysics is often overlooked, and restricted to only the spectral cross-talk term, i.e. the excitation of the red fluorescent dyes by the green laser. It is therefore widely assumed that the fluorescence emission of the red dye (the acceptor) under red excitation is not affected by the presence of the green laser excitation, especially in the case of alternating laser excitations (ALEX and PIE-FRET). Quite surprisingly, we show here that this assumption turns out to be inaccurate and that the presence of the green laser light can have a non-negligible influence on the emission of common red fluorescent dyes Alexa Fluor 647 and Atto 647N. Even in the case where the alternating green and red laser pulses are temporaly delayed by a time much longer than the fluorophore's excited state lifetime and time gating is used to select only the fluorescence excited by the red laser, we observe that the emission of the red fluorescent molecule can be significantly reduced by the presence of the green laser. This effect is important to account for the accurate distance determination in single-molecule FRET experiments as it affects the acceptor brightness [10]. Likewise, this phenomenon is also important for the accurate concentration determination in FCCS analysis. The fact that we obtain similar observations for a cyanine dye (Alexa Fluor 647, a Cy5 analog) and a carbo-rhodamine dye (Atto 647N) tends to indicate that this phenomenon is quite general [13]. Hereafter, we will quantify its influence and provide guidelines so as to avoid introducing experimental artifacts in PIE-FRET and FCCS measurements.

## 2. Methods

Alexa Fluor 647 (Thermo Fisher Scientific) and Atto 647N (AttoTec) are used as received from the manufacturer and diluted to nanomolar concentrations in phosphate-buffered saline (PBS) solution (Sigma Aldrich). The confocal microscope is custom built to perform PIE-FRET with alternating green and red picosecond laser pulses at 40 MHz repetition rate [12]. The green laser pulses are provided by a iChrome-TVIS laser (Toptica) at 557 nm wavelength (pulse duration ~ 3 ps), while the red laser pulses stem from a LDH-P-635 laser diode (Picoquant) at 635 nm (pulse duration < 70 ps). Both lasers are synchronized to the same 40 MHz clock and temporarily delayed respective to each other by 12.5 ns, which is half the repetition period. This realizes a pulse train of alternating colors at 557 and 635 nm. The 557 nm laser is modulated in the millisecond timescale by a rotating optical chopper blade (Thorlabs MC1F10) set at the focus position of the telescope expanding system, where the lateral beam size is minimum.

The laser beams are spatially overlaped and reflected towards the microscope by a multi-band dichroic mirror (ZT405/488/561/647rpc, Chroma Technology Corp). Focusing is performed by a high numerical aperture water-immersion objective (C-Apochromat 63x 1.2NA, Zeiss) approximately 15 µm inside the solution containing fluorescent molecules. The fluorescence light is collected by the same objective and directed towards a 50 µm pinhole conjugated to the object plane. Photon counting is performed after the confocal pinhole by two avalanche photodiodes (MPD-5CTC, Picoquant) separated by a 50/50 non-polarizing cube beamsplitter (Melles Griot). Back-scattered laser light is filtered by 670 ± 20 nm bandpass filters (FF01-676/37, Semrock). The photodiode signals are recorded by a fast time-correlated single photon counting module (Hydraharp400, Picoquant) in time-tagged time-resolved (TTTR) mode. Fluorescence spectra are recorded after the confocal pinhole by a QE65-Pro spectrophotometer (Ocean Optics). Calibration of the microscope point spread function using individual 100 nm fluorescent beads (Tetraspec, Thermo Fisher Scientific) indicates a confocal volume of 0.8 femtoliter.

Since the green and red laser pulses are separated a 12.5 ns delay which is much longer than the 1 ns fluorescence lifetime of Alexa 647 (or 3.5 ns for Atto 647N), the fluorescence steming from the green or red excitation can be readily distinguished by time gated analysis. Hereafter, we select only the fluorescence photons originating from the excitation by the red laser by applying a time gate on the fluorescence data from 13 to 25 ns (Figure 1a). All fluorescence time traces are analyzed using Symphotime64 software (Picoquant). Fitting the FCS correlation data follows a three dimensional Brownian diffusion model with an additional blinking term [14]:

$$G(\tau) = \frac{1}{N}\left(1 + \frac{T}{1-T}\exp\left(-\frac{\tau}{\tau_T}\right)\right)\frac{1}{(1+\frac{\tau}{\tau_d})\sqrt{1+\frac{1}{\kappa^2}\frac{\tau}{\tau_d}}} \quad (1)$$

where N is the total number of molecules, T the fraction of dyes in the dark state, $\tau_T$ the dark state blinking time, $\tau_d$ the mean diffusion time and κ the aspect ratio of the axial to transversal dimensions of the confocal volume which is fixed to κ = 5 for the FCS analysis.

3. Results and Discussion

We start by monitoring the fluorescence intensity from Alexa Fluor 647 molecules (Figure 1a,b). Time gating of the detection events on the 13-25 ns temporal window centered on the arrival time of the red excitation laser pulse discriminates the fluorescence excited by the 635 nm laser from the one excited by the 557 nm light. To better present the results, hereafter we select the photon detection events corresponding to Alexa 647 fluorescence excited by the red laser at 635 nm. Since the green and red laser pulses are delayed by a time much greater than the 1 ns fluorescence lifetime of Alexa 647 and since only the fluorescence excited by the red laser is kept for analysis, one would expect that the green laser has no effect on Alexa 647 fluorescence. Alternatively, as more total energy is brought onto the sample, one could expect that the total fluorescence signal increases due to the direct excitation of Alexa 647 by the 557 nm laser light which sums up with the fluorescence excited by the 635 nm laser. However, quite surprisingly, the fluorescence intensity *decreases* by about 20% when the 557 nm laser illuminates the sample (Figure 1b). This phenomenon is fully reversible by switching off the green laser, and prolongated observations for more than 20 minutes showed no obvious signs of photobleaching or signal decrease. If no gating is applied on the detection, a drop of fluorescence intensity is still observed with the presence of the green laser light. The drop amplitude is reduced from 20 to 10% (due to the additional contribution of the fluorescence excited by 557 nm photons), but it remains visible. To the best of our knowledge, this is the first report of red fluorescence quenching induced by a time delayed green laser light.

To decrease the Alexa 647 fluorescence intensity, the green excitation must affect the dye photodynamics on a timescale longer than the 12.5 ns delay between the green and red laser pulses. We therefore investigate Alexa 647 photodynamics by reconstructing the photon arrival time histograms with time correlated single photon counting (TCSPC, Figure 1c) and by computing the fluorescence intensity temporal correlation (FCS, Figure 1d). The TCSPC fluorescence lifetime decay traces in Figure 1c display a drop in amplitude by 20% upon illumination by the green laser, consistent with the fluorescence intensity data in Figure 1b. However, the normalized decay traces (insert Figure

1c) nearly perfectly overlap, indicating that the 1 ns fluorescence lifetime of Alexa 647 is unchanged by the 557 nm illumination.

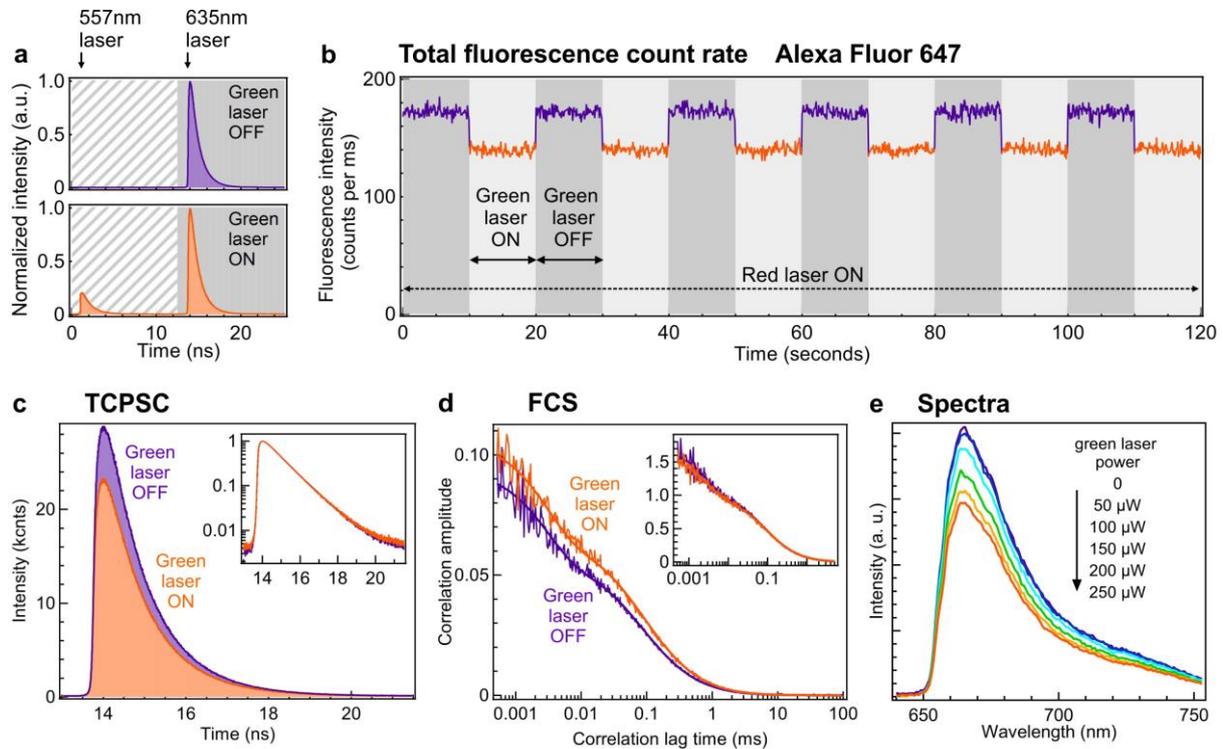

**Figure 1.** Alexa Fluor 647 emission modification upon illumination at 557 nm. (a) Scheme of the PIE-FRET excitation with alternating picosecond laser pulses at 557 nm and 635 nm. The green and red laser pulses are temporaly delayed by 12.5 ns with respect to each other, this corresponds to half a period of the 40 MHz repetition rate. Only the time-gated fluorescence excited by the red laser beam is used for the analysis (photon time interval 12.5 – 25 ns, grey shadowed area). (b) Fluorescence intensity time trace of Alexa Fluor 647 molecules with periods when the 557 nm laser is switched on and off (557 nm laser power 230 µW, 635 nm laser power 45 µW). Despite the fact that the green and red excitation laser pulses are separated by a time much longer than Alexa 647 lifetime, the 557 nm green excitation induces a significant reduction of Alexa 647 fluorescence intensity. (c) Lifetime decay data of the fluorescence time traces in (b), the insert displays the normalized decay traces in logarithmic scale. (d) Fluorescence correlation spectroscopy data (thin lines) and numerical fits (thick lines) of the fluorescence time traces in (b). The insert displays the normalized FCS correlograms. Results from the FCS analysis are summarized in Table 1. (e) Fluorescence spectra for increasing values of 557 nm laser power, taken at 100 µW red laser power. The spectra are recorded after the confocal pinhole to ensure the signal is comparable to the time trace data.

FCS analysis investigates the temporal dynamics occurring on the microsecond to millisecond timescale. As a Cy5 analog, Alexa 647 is known to feature cis-trans photoisomerization inducing apparent blinking on the intensity time trace with typical blinking time of a few microseconds [15,16]. This phenomenon is visible on the FCS data by a shoulder on the correlogram at microsecond timescales (Figure 1d and 2a) in addition to the main correlation contribution around 0.1 ms induced by translational diffusion across the confocal volume. The major effect induced by the presence of the green laser is a net increase of the correlation amplitude which corresponds to a reduction in the observed number of fluorescent Alexa 647 molecules in the confocal volume. Table 1 summarizes the results of FCS data fitting with the model in Eq. (1). The main influence of the green excitation light is a decrease of the number of apparent fluorescent-active molecules from N = 18 to 15 in the presence of the green laser light (Figure 2b). The fluorescence brightness per molecule (estimated by dividing the mean fluorescence intensity by the average number of molecules measured by FCS) is also reduced upon 557 nm illumination from 9.6 to 9.3 counts per ms (Figure 2e). The other FCS observables (diffusion time, blinking time, dark state fraction) remain unchanged within experimental uncertainties (Figure 2c and 2d). The fact that the translationnal diffusion time $\tau_d$ is unchanged is another confirmation that photobleaching is not affecting our observations since photobleaching would have the consequence to reduce $\tau_d$ as the green power is increased [16,17]. Likewise, saturation of fluorescence phodynamics is known to induce an increase in the detected number of molecules and their mean diffusion time accross the confocal volume [17], two features that are absent in our observations.

**Table 1.** Results of the FCS analysis for the data displayed in Fig. 1d following the model of Eq. (1).

|  | Green laser OFF | Green laser ON |
|---|---|---|
| N | 18.0 ± 0.7 | 15.0 ± 0.7 |
| $\tau_d$ (µs) | 87 ± 5 | 92 ± 5 |
| T | 0.40 ± 0.02 | 0.39 ± 0.02 |
| $\tau_T$ (µs) | 3 ± 1 | 3 ± 1 |
| Brightness per molecule (kHz) | 9.6 ± 0.3 | 9.3 ± 0.3 |

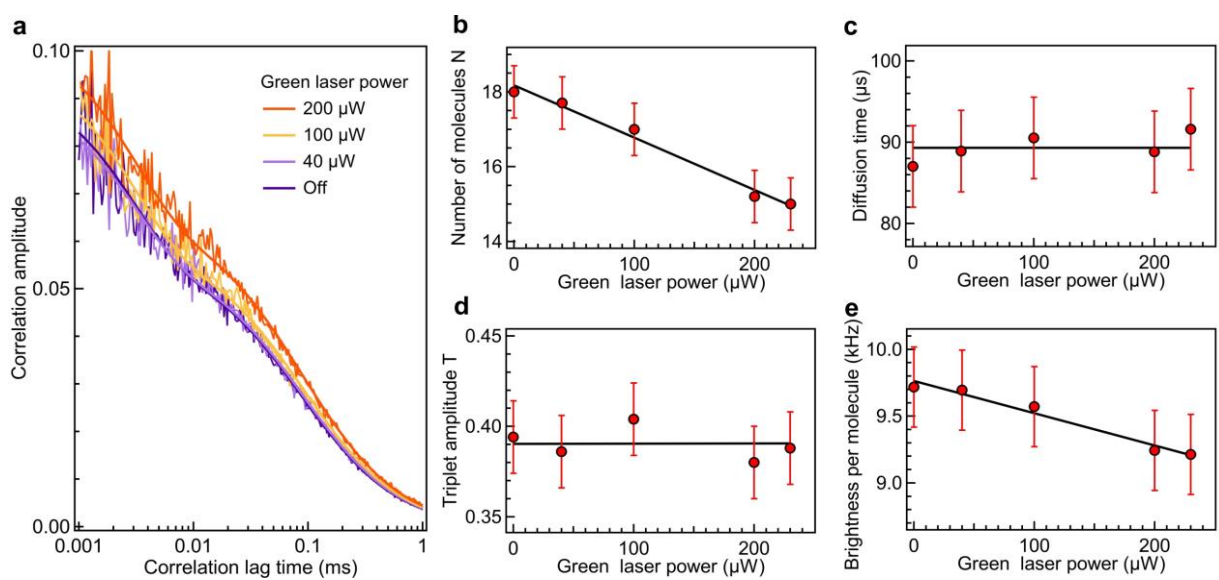

**Figure 2.** Evolution of FCS data on Alexa 647 with the green excitation power at 557 nm. (a) FCS data (thin lines) and numerical fits (thick lines) recorded for increasing powers of the green excitation. The results from the FCS analysis are summarized in (b-e), with the number of molecules N (b), the diffusion time $\tau_d$ (c), the triplet amplitude T (d) and the fluorescence brightness per molecule (e). Lines are empirical fits to guide the eyes. Like the triplet amplitude and the diffusion time, the blinking time $\tau_T$ was found constant within the measurement uncertainties.

To complete the data recorded on Alexa 647, the fluorescence spectrum is monitored (Figure 1e). This data is recorded after the confocal pinhole to ensure that the fluorescence signal stems from a similar configuration (excitation, collection) than for the fluorescence time trace. Again, a decrease of the total intensity is observed as the 557 nm power increases, in agreement with the data shown in Figure 1b. After intensity normalization, the different spectra nearly perfectly overlap so that we can rule out any spectral change of Alexa 647 emission upon 557 nm illumination.

While FCS can monitor fluorescence intensity changes occurring in the μs to ms time range, it is however limited to millisecond time scale for the case of freely diffusing fluorescent dyes. To investigate the fluorescence photodynamics occurring at a timescale longer than a few milliseconds, we modulate the 557 nm laser intensity by an optical chopper blade, switching the laser intensity on and off at a given frequency in the 10-1000 Hz range. For low modulation frequencies (below 50 Hz), the fluorescence intensity follows an exponential recovery after the green laser is switched off (Fig. 3a). This is characteristic of a long-lived dark state whose lifetime is equivalent to the exponential recovery time. In other words, the dark state populated by the 557 nm laser requires a few millisecond for its population to go down to the ground state. Fitting the data in Fig. 3a,b yields an average dark state lifetime of 3.8 ± 1.2 ms. When the modulation frequency is increased over 100 Hz, the exponential recovery feature disappears as the half-period becomes shorter than the characteristic exponential time. This provides another confirmation that the dark state lifetime is on the order of 4 ms.

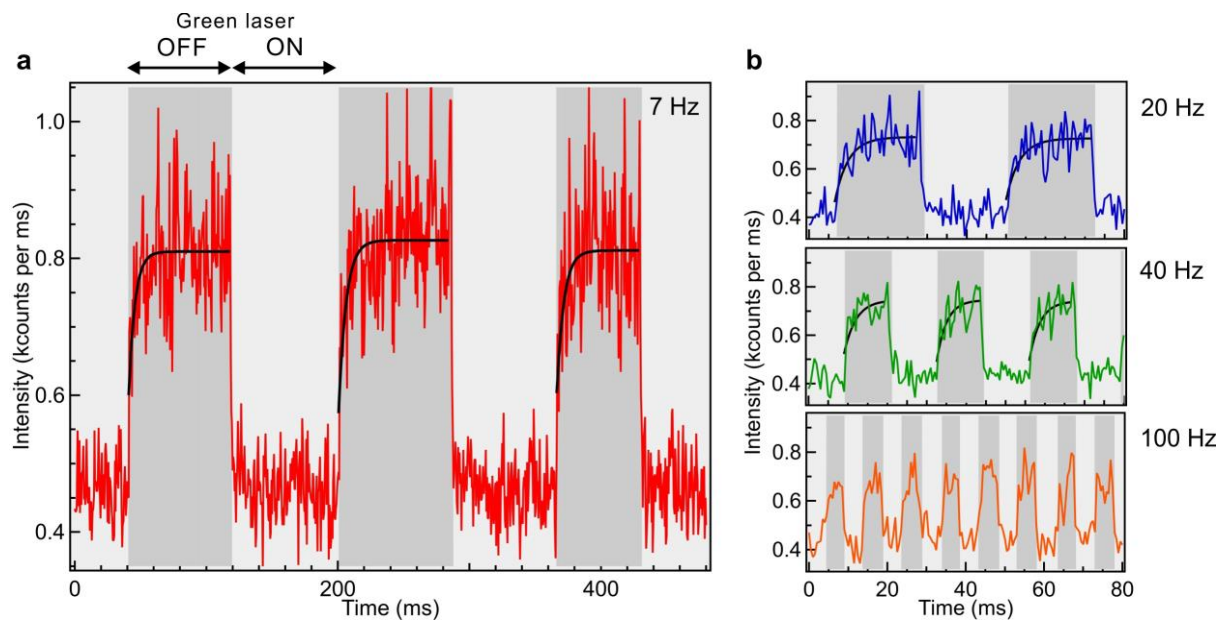

**Figure 3.** Millisecond timescale dynamics of Alexa 647 fluorescence upon modulated excitation at 557 nm. (a) Fluorescence intensity time trace at 7 Hz modulation of the 557 nm laser. After the green laser is switched off, an exponential recovery with ~ 4 ms characteristic time is observed (black traces are numerical fits). (b) Fluorescence time traces for increasing modulation frequencies. The exponential recovery feature is barely noticeable at 100 Hz modulation while the fluorescence modulation amplitude (difference between min and max levels) also vanishes. For all the data shown here, the 557 nm laser power was 230 µW, the 635 nm laser power was 45 µW and the binning time was 0.5 ms.

Finally, we investigate the power dependence of Alexa 647 fluorescence intensity so as to provide guidelines on how to avoid this quenching phenomenon by adjusting the laser powers accordingly. Figure 4a summarizes the results for increasing values of red and green laser powers. The effects of fluorescence saturation become visible for 635 nm laser powers exceeding 200 µW, and become even more pronounced as the 557 nm laser power is increased. To analyze this data, we model the detected fluorescence intensity $F$ by [18]:

$$F = A \frac{I_{635}}{1+\frac{I_{635}}{I_{sat}}} \quad (2)$$

where $I_{635}$ is the 635 nm average laser power, $I_{sat}$ the saturation intensity and *A* is a proportionality factor. With increasing 557 nm power, we monitor a decrease on both the amplitude *A* and the saturation intensity $I_{sat}$ (Figure 4b,c). This characterization provides important guidelines so as to set the typical laser powers in order to avoid entering into the saturation regime of the fluorescence photophysics.

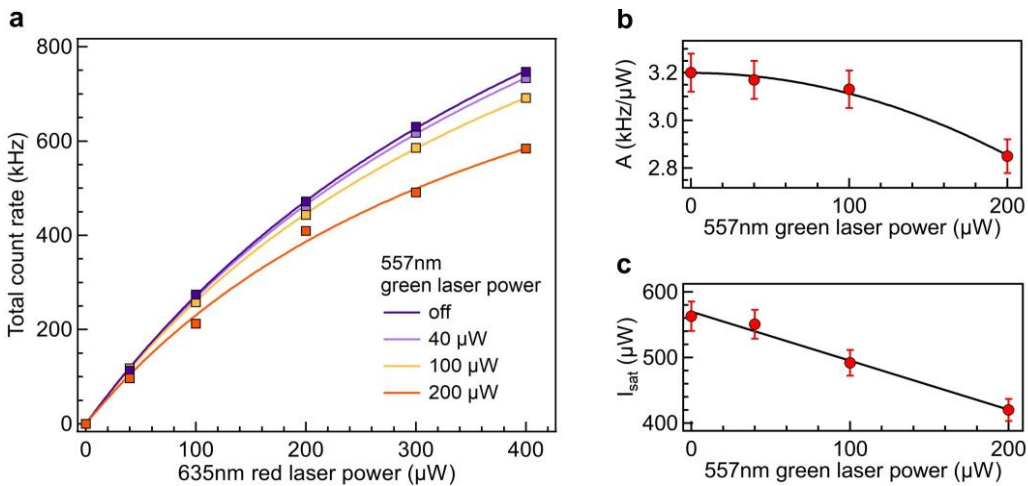

**Figure 4.** Excitation power dependence of Alexa 647 fluorescence intensity. (a) Evolution of the total detected fluorescence intensity as a function of the 635 nm laser power and for increasing values of 557 nm laser power. The lines are numerical fits following the model in Eq. (2), which are used to extract the parameters A and $I_{sat}$ presented in (b) and (c) as functions of the green laser power. Black lines in (b) and (c) are guide to the eyes.

All the aforementioned experimental results were recorded for Alexa 647. To check whever our observations are related to the cyanine structure of the dye, we perform similar experiments on Atto 647N, which features a carbo-rhodamine structure (Figure 5a) [13]. Interestingly, we monitor a similar drop of Atto 647N fluorescence intensity upon extra illumination by 557 nm laser pulses (Figure 5b,c). These observations tend to indicate that the quenching phenomenon induced by the green excitation is quite general and not related to the dye structure. However, some noticeable changes are visible between Alexa 647 and Atto 647N behaviours. First, the saturation intensity $I_{sat}$ in the case of Atto 647N is found to be independent of the green excitation power, contrarily to the Alexa 647 case (Figure 3c). Moreover, detailed FCS analysis for Alexa 647 and Atto 647N indicate additional features (Figure 5d-f). Both Alexa 647 and Atto 647N experience similar relative decrease in their net fluorescence intensity upon 557 nm illumination (Figure 5d). For both dyes, FCS monitors a decrease in the apparent number of fluorescent molecules $N$ in the confocal volume which scales linearly with the green excitation power (Figure 5e). However, the drop in $N$ appears more pronounced for Atto 647N than for Alexa 647. Lastly, the brightness per molecule decreases for Alexa 647 while the 557 nm power is increased, while for Atto 647N, we find that the fluorescence brightness per molecule is mostly independent of the green excitation power.

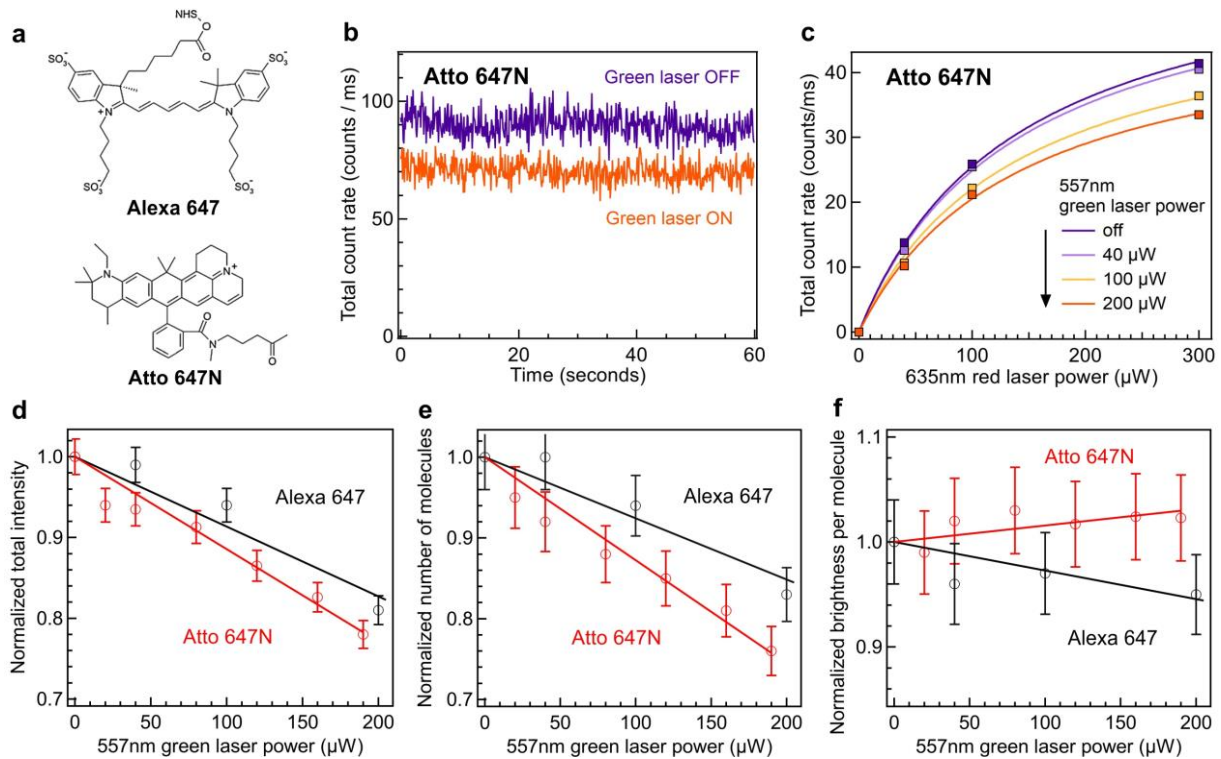

**Figure 5.** Comparison between Alexa 647 and Atto 647N. (a) Molecular structures of the cyanine Alexa Fluor 647 and the carbo-rhodamine Atto 647N. (b) Fluorescence intensity time trace recorded for Atto 647N with 40 µW laser power at 635 nm with or without 200 µW laser power at 557 nm. (c) Excitation power dependence of Atto 647N fluorescence intensity for different values of green laser power. Following FCS analysis and normalization to the case when the green laser is switched off, the results for Alexa 647 and Atto 647N can be compared for the total fluorescence intensity (d), the detected number of molecules (e) and the brightness per molecule (f) as a function of the green laser power. Lines are empirical fits to guide the eyes.

Altogether, these experimental data provide the picture that the 557 nm excitation drives the red fluorescent molecules into a long-lived dark state, hereby reducing the apparent number of red fluorescent-active molecules in the detection volume and lowering the detected fluorescence intensity. Since the green and red laser pulses are delayed by a time much longer than the fluorophore's excited state lifetime, the excitation of the dark state by the green laser is likely to start from the fluorophore's ground state. Moreover, since the delay between green and red excitation pulses is 12.5 ns, the typical residence time of the dark state driven by the green excitation must exceed 12 ns. As FCS analysis did not indicate any additional blinking on the timescale from 0.5 µs to 1 ms, we expect that the dark state typical residence time is greater than one millisecond. This conclusion is supported by the experimental results where the green power is temporaly modulated in the 10-100 Hz range (Fig. 3). This relatively long lifetime in the millisecond range appears typical of long-lived triplet state which are populated by inter system crossing from the excited state or from higher excited states. There is also the possibility of multiple absorption events [17], where molecules excited by the green laser further absord energy from the red laser before entering into the dark state.

Another possible explanation would involve the contribution of photorefractive effects induced by the green laser, i.e. the green laser affecting the focusing and fluorescence collection for the red fluorescence. Although the optical powers involved remain a at low level on the order of a hundred of microwatts and photorefractive effects are expected to be negligible, we presently cannot rule out this possibility. Altogether, these observations are important to avoid introducing experimental artifacts in PIE-FRET, ALEX-FRET [10–12] or FCS / FCCS analysis [19], since both the apparent number of fluorescent active molecules and their brightness can be affected.

## 4. Conclusion

Here we have demonstrated that the presence of green laser pulses can quench the fluorescence of common red dyes Alexa Fluor 647 and Atto 647N, despite the fact that the green laser pulses are temporaly delayed by a time much greater than the fluorophore lifetime. These are important observations as multi-color approaches to single fluorescent molecules (PIE-FRET, ALEX-FRET, FCCS…) are receiving a growing interest. In order to avoid introducing experimental artifacts affecting the fluorescence brightness and apparent fluorophore concentrations, it is advisable to keep the average powers below 40 µW for both red and green lasers at 40 MHz repetition rate. The physical origin of this phenomenon could be linked to a long-lived dark state of the red dyes on the millisecond time scale driven by the 557 nm excitation. Additionally, the green laser could induce photorefractive effects in the microscope objective, reducing the apparent size of the confocal detection volume and lowering the collection efficiency. Lastly, as our observations are found to be highly dependent on the spatial overlap between the green and red laser beam, this phenomenon can also be used as an additional feature for aligning the multicolor microscope: observing a drop of red fluorescence intensity in the presence of the green laser is the sign that both laser beams are correctly aligned and overlaped.

5. Acknowledgments

This project has received funding from the Agence Nationale de la Recherche (ANR) under grant agreement ANR-17-CE09-0026-01.